\documentclass[12pt]{iopart}
\usepackage[T1]{fontenc}
\usepackage[ascii]{inputenc}
\usepackage{cite}
\usepackage{graphicx}
\usepackage{microtype}
\usepackage{iopams}
\usepackage{hyperref}
\newcommand{\imag}{\mathrm{Im}\,}

\newcommand{\PT}{\mathcal{PT}}

\newcommand{\mrm}{\mathrm}
\newcommand{\rmk}{\mathrm{k}}
\newcommand{\rmj}{\mathrm{j}}

\makeatletter
\newcommand{\eqlabel}[1]{{%
  \edef\@currentlabel{\arabic{eqnval}}%
  \label{#1}}}
  \addtocounter{eqnval}{1}
\makeatother

\begin{document}

\title[Eigenvalue structure of a BEC in a $\PT$-symmetric double well]%
{Eigenvalue structure of a Bose-Einstein condensate in a $\PT$-symmetric
double well}

\author{Dennis Dast, Daniel Haag, Holger Cartarius, J\"org Main and
G\"unter Wunner}

\address{1. Institut f\"ur Theoretische Physik, Universit\"at Stuttgart,
70550 Stuttgart, Germany}
\eads{\mailto{Dennis.Dast@itp1.uni-stuttgart.de},
\mailto{Daniel.Haag@itp1.uni-stuttgart.de}}
\begin{abstract}
We study a Bose-Einstein condensate in a $\PT$-symmetric double-well potential
where particles are coherently injected in one well and removed from the other
well. In mean-field approximation the condensate is described by the
Gross-Pitaevskii equation thus falling into the category of nonlinear
non-Hermitian quantum systems. After extending the concept of $\PT$ symmetry
to such systems, we apply an analytic continuation to the Gross-Pitaevskii
equation from complex to bicomplex numbers and show a thorough numerical
investigation of the four-dimensional bicomplex eigenvalue spectrum. The
continuation introduces additional symmetries to the system which are confirmed
by the numerical calculations and furthermore allows us to analyze the
bifurcation scenarios and exceptional points of the system. We present a linear
matrix model and show the excellent agreement with our numerical results. The
matrix model includes both exceptional points found in the double-well
potential, namely an EP2 at the tangent bifurcation and an EP3 at the pitchfork
bifurcation. When the two bifurcation points coincide the matrix model
possesses four degenerate eigenvectors. Close to that point we observe the
characteristic features of four interacting modes in both the matrix model and
the numerical calculations, which provides clear evidence for the existence of
an EP4.
\end{abstract}

\maketitle

\section{Introduction}
Non-Hermitian Hamiltonians obeying parity-time ($\PT$) symmetry can have
entirely real eigenvalues, which is referred to as the case of unbroken $\PT$
symmetry~\cite{Bender98a}. Much effort has been put into the development of a
more general formulation of quantum mechanics, where the requirement for
Hermiticity is replaced by the weaker requirement of $\PT$
symmetry~\cite{Bender07a,Bender02a,Bender99a,Mostafazadeh08a,Mostafazadeh10a}.
This includes the construction of a unitary time development and a positive
definite scalar product in the case of unbroken $\PT$
symmetry~\cite{Bender07a,Mostafazadeh07a}. Also it is possible to embed $\PT$
symmetry in the more general approach of
pseudo-Hermiticity~\cite{Mostafazadeh02a,Mostafazadeh02b,Mostafazadeh02c}.

$\PT$ symmetry is not only the subject of theoretical but also of experimental
studies~\cite{Rueter10a,Guo09a,Schindler11a,Bittner12a}. In particular the
observation of $\PT$ symmetry in optical wave guide systems was a great
success~\cite{Klaiman08a,Rueter10a,Guo09a}. The experimental realization in a
true quantum system, however, is still missing although suggestions already
exist~\cite{Kreibich13a} for systems which have been extensively
studied~\cite{Elsen11a,Rapedius10a}. In contrast to the above-named
generalization of quantum mechanics, in experimental realizations one acts on
the assumption that the complete system is Hermitian. Here the non-Hermiticity
enters the picture as an effective description of an open quantum system by
using it to model gain and loss contributions. The system we will investigate
falls into this category. Following a suggestion by Klaiman
\etal~\cite{Klaiman08a} we study a Bose-Einstein condensate (BEC) where
particles are coherently injected and removed from the system. Both a coherent
influx and outflux of atoms have been experimentally realized. Electron beams
have been successfully used to ionize atoms of the BEC and extract them with an
electrostatic field~\cite{Gericke08a}. Coherent incoupling of atoms into a BEC
from an additional source condensate has been realized by exploiting various
electronic excitations of the incoupled atoms~\cite{Robins08a}.

While most of the work done in the field of $\PT$ symmetry treats the linear
Schr\"odinger equation, a BEC in mean-field approximation is described by the
nonlinear Gross-Pitaevskii equation (GPE). It is known that this equation
provides accurate results for temperatures considerably smaller than the
critical temperature of the condensate~\cite{Gross61a,Pitaevskii61a}, however,
for dynamically unstable states the mean-field description breaks
down~\cite{Vardi01a,Trimborn09a,Fallani04a}. The nonlinearity leads to
remarkable effects, such as the coexistence of $\PT$-symmetric and $\PT$-broken
states in certain parameter
regimes~\cite{Cartarius12a,Cartarius12b,Dast12a,Graefe12b}. 

\begin{figure}%
  \centering%
  \includegraphics[width=0.8\columnwidth]{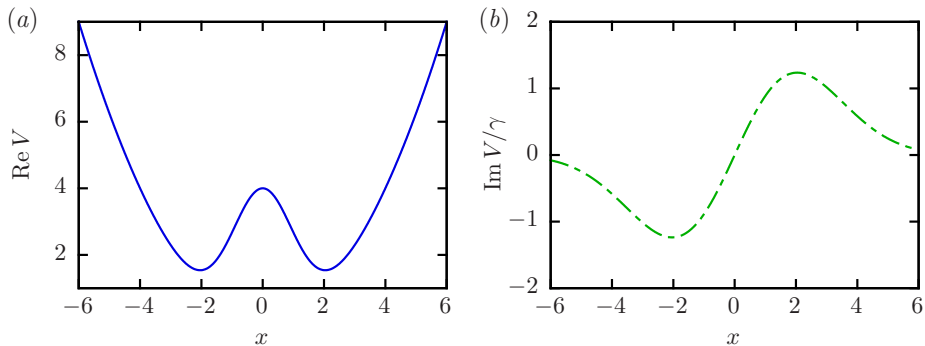}%
  \caption{The one-dimensional $\PT$-symmetric double-well potential $V$. The
    real part is a symmetric and the imaginary part an antisymmetric function.
    The imaginary part describes a gain (loss) in the right (left) well.}%
  \label{fig:potential}%
\end{figure}

In this paper we analyse the spectral properties of a BEC with contact
interaction described by the dimensionless GPE,
\begin{equation}
  \left[-\frac{\partial^2}{\partial x^2} +V(x)
    -g\left|\psi(x,t)\right|^2\right] \psi(x,t)
    = \rmi \frac{\partial}{\partial t}\psi(x,t),
  \label{eq:gpe}
\end{equation}
in the $\PT$-symmetric double-well potential
\begin{equation}
  V(x)=\frac14 x^2 + v_0 \rme^{-\sigma x^2} + \rmi\gamma x\rme^{-\rho x^2},
  \label{eq:potential}
\end{equation}
shown in \fref{fig:potential}. Imaginary potentials are sinks or sources of
probability density depending on the sign of the imaginary part. The physical
interpretation is a coherent injection or removal of particles into or from the
condensate. The non-Hermitian GPE can also be obtained in the mean-field limit
of an open Bose-Hubbard model described by a master
equation~\cite{Trimborn08a,Witthaut11a}. The symmetric real part of the
potential forms a double well consisting of a harmonic trap and a Gaussian
barrier with height $v_0$ and width parameter $\sigma$. The parameter $\rho$ of
the antisymmetric imaginary part is chosen in such a way that its extrema
coincide with the minima of the double well. Positive values of $g$ describe an
attractive contact interaction whereas negative values relate to a repulsive
interaction. The separation $\psi(x,t)=\rme^{-\rmi\mu t}\psi(x)$ leads to the
time-independent GPE with the (nonlinear) eigenvalue $\mu$ which can be
identified with the chemical potential and the (nonlinear) eigenstate
$\psi(x)$. In all following calculations $v_0=4$, $\sigma=0.5$ and
$\rho\approx0.12$ are fixed, while the gain/loss parameter $\gamma$ and the
strength of the nonlinearity $g$ are varied to study the behaviour of the
system. For the sake of clarity only a one-dimensional system is considered. An
extension to three dimensions with additional harmonic potentials in $y$ and
$z$ direction does not show further effects~\cite{Dast12a}.

Exact numerical results are gained by integrating the one-dimensional wave
function outwards and by fulfilling the boundary conditions via a root search.

Before studying a BEC in the $\PT$-symmetric double well it is necessary to
expand the concept of linear $\PT$ symmetry to nonlinear systems, which will be
done in \sref{sec:pt}. We apply an analytic continuation to the GPE from
complex to bicomplex numbers in \sref{sec:spectrum} and present a full analysis
of the four-dimensional bicomplex eigenvalue spectrum, the eigenfunctions and
the symmetries of the system. The continuation allows the detailed
investigation of the exceptional points of the system and their mathematical
structure in \sref{sec:ep}. A comparison of the eigenvalues and the exceptional
points with those of a linear $4\times4$ matrix model is drawn in
\sref{sec:matrixmodel}. Finally it is shown in \sref{sec:EP4} that the matrix
model has the interesting feature of four interacting eigenmodes in the limit
of vanishing interaction, and that this behaviour can also be found in the GPE.

\section{$\PT$ symmetry in nonlinear systems}
\label{sec:pt}
A linear system is called $\PT$-symmetric if its Hamiltonian commutes with the
combined action of the parity and time reversal operator $[\hat{H}, \PT]=0$.
With $\hat{H}=\hat{p}^2/2m+V(\hat{x})$ and the operators $\mathcal{P}:\
\hat{x}\rightarrow-\hat{x},\ \hat{p}\rightarrow-\hat{p}$ and $\mathcal{T}:\
\hat{p}\rightarrow-\hat{p},\ \rmi\rightarrow-\rmi$ one gains the necessary
condition $V(\hat{x})=V^*(-\hat{x})$, i.e.\ the real part of the potential must
be symmetric and the imaginary part antisymmetric. Linear $\PT$-symmetric
systems have the following properties~\cite{Bender07a}:

If an eigenstate of $\hat{H}$ is also an eigenstate of the operator $\PT$,
i.e.\ $\PT\psi=\exp(\rmi\varphi)\psi,\ \varphi\in\mathbb{R}$, then the
corresponding energy eigenvalue $E$ is real. Also the inverse statement holds.
The eigenspace to a real eigenvalue can always be constructed out of
$\PT$-symmetric states.

With these two properties the following important conclusion can be derived.
If and only if all eigenstates of the Hamiltonian can be written as eigenstates
of the $\PT$ operator, the spectrum is entirely real. One refers to this case
as unbroken $\PT$ symmetry, otherwise the $\PT$ symmetry is broken. Complex
eigenvalues appear as complex conjugate pairs and their wave functions can be
mapped on each other by application of the $\PT$ operator.

In nonlinear systems like BECs, we cannot use the simple commutator relation
$[\hat{H},\PT]=0$ to define $\PT$ symmetry. In the following we expand this
condition to nonlinear systems described by a Gross-Pitaevskii-like equation,
\begin{equation}
  \hat{H}_\mrm{lin} \psi + f\left(\psi\right)\psi
    =\rmi \frac{\partial}{\partial t}\psi,
  \label{eq:nonlinear_SGL}
\end{equation}
where $\hat{H}_\mrm{lin}=-\hat{p}^2/2m+V(\hat{x})$ and $f\left(\psi\right)$ is
a general nonlinear part. We restrict the discussion to nonlinearities which
are invariant under the change of a global phase,
\begin{equation}
  f\left(\rme^{\rmi\chi}\psi\right)=f\left(\psi\right), \quad\chi\in\mathbb{R}.
  \label{eq:f_cond1}
\end{equation}
This is necessary for the time-independent nonlinear Schr\"odinger equation to
have the usual form. For nonlinear systems we replace the commutator relation
of linear $\PT$-symmetric systems with the requirement
\begin{equation}
  \PT \left[\left(\hat{H}_\mrm{lin} + f\left(\psi\right)\right)\psi\right]
    \stackrel{!}{=}
    \left[\hat{H}_\mrm{lin} + f\left(\PT\psi\right)\right]\PT\psi.
  \label{eq:nonlinear_commutator}
\end{equation}
We will see that this condition suffices to regain the properties of linear
$\PT$ symmetry. If the linear part $\hat{H}_\mrm{lin}$ is $\PT$-symmetric,
\eref{eq:nonlinear_commutator} is reduced to a simple condition for the
nonlinear part $f(\psi)$,
\begin{equation}
  \PT f\left(\psi\right) = f\left(\PT\psi\right).
  \label{eq:f_cond2}
\end{equation}

In this work we are mainly interested in the properties of stationary states.
Therefore we look at the time-independent nonlinear Schr\"odinger equation,
\begin{equation}
  \hat{H}_\mrm{lin} \psi + f\left(\psi\right)\psi = \mu \psi.
  \label{eq:nonlin_stat}
\end{equation}
Application of the $\PT$ operator leads to
\begin{equation}
  \hat{H}_\mrm{lin} \PT\psi + f\left(\PT\psi\right)\PT\psi = \mu^*\PT\psi,
  \label{eq:PT_nonlin_stat}
\end{equation}
where \eref{eq:f_cond2} and $[\hat{H}_\mrm{lin},\PT]=0$ was used.
Equations~\eref{eq:nonlin_stat} and \eref{eq:PT_nonlin_stat} show immediately
that the energy eigenvalues $\mu$ occur in complex conjugate pairs with the
eigenstates $\psi$ and $\PT\psi$, respectively. Also the most striking property
of $\PT$ symmetry, namely the concurrence of $\PT$-symmetric states and real
eigenvalues, is true for such nonlinear systems. This can be seen by evaluating
\eref{eq:PT_nonlin_stat} for $\PT$-symmetric states
$\PT\psi=\exp(\rmi\varphi)\psi$,
\begin{equation}
  \hat{H}_\mrm{lin}\psi + f\left(\rme^{\rmi\varphi}\psi\right)\psi = \mu^*\psi.
  \label{eq:nonlinear_pt_symmetric}
\end{equation}
As stated in \eref{eq:f_cond1} only phase independent nonlinearities are
considered, and thus for $\PT$-symmetric states the energy eigenvalue must be
real $\mu=\mu^*$.

Again this proof is also valid in the inverse direction. For non-degenerate
real eigenvalues \eref{eq:nonlin_stat} and \eref{eq:PT_nonlin_stat} demand that
the eigenfunction is $\PT$-symmetric because $\psi$ and $\PT\psi$ fulfil the
same eigenvalue equation. However if a real eigenvalue is degenerate, it is in
general not possible to choose $\PT$-symmetric eigenstates because the
superposition principle is only valid in linear systems.

These results can be summarized as follows. In nonlinear non-degenerate systems
of type \eref{eq:nonlinear_SGL} with a $\PT$-symmetric linear part and a
nonlinear part which fulfils the conditions \eref{eq:f_cond1} and
\eref{eq:f_cond2}
\begin{itemize}
  \item the eigenvalues are either real or occur in complex conjugate pairs,
  \item the eigenvalues are real if and only if the eigenstate itself is
    $\PT$-symmetric,
  \item if $\psi$ is an eigenstate to $\mu$ then $\PT\psi$ is eigenstate to
    $\mu^*$.
\end{itemize}
It is worth mentioning that for $\PT$-symmetric wave functions $\psi$ the
conditions \eref{eq:f_cond1} and \eref{eq:f_cond2} also guarantee that
$f(\psi)$ is $\PT$-symmetric, i.e.\ the real part of $f$ is symmetric and the
imaginary part is antisymmetric. Since for a given state $\psi$ the
nonlinearity has to be seen as a contribution to the potential $V$, this
establishes the link to linear $\PT$ symmetry.

These general considerations are now applied to the Gross-Pitaevskii
nonlinearity,
\begin{equation}
  f(\psi,\bi{r}) = \int\rmd^3\bi{r}'\, V\left(\bi{r},\bi{r}'\right) 
      \left|\psi(\bi{r}')\right|^2.
  \label{eq:GPE_nonlin}
\end{equation}
Because the wave function only appears as square modulus, the nonlinearity is
not changed by an arbitrary phase and thus condition~\eref{eq:f_cond1} is
always fulfilled independent of the interaction type. The second
condition~\eref{eq:f_cond2} carries over to a condition for the interaction
potential $V(\bi{r},\bi{r}')$,
\begin{eqnarray}
  \int\rmd^3\bi{r}'\, V^*\left(-\bi{r},\bi{r}'\right) 
    \left|\psi(\bi{r}')\right|^2 =
    \int \rmd^3 \bi{r}'\, V\left(\bi{r},\bi{r}'\right) 
    \left|\psi^*(-\bi{r}')\right|^2,\nonumber\\
    V\left(\bi{r},\bi{r}'\right) =
    V^*\left(-\bi{r},-\bi{r}'\right).
  \label{eq:int_pot_condition}
\end{eqnarray}
The most common interaction potentials, namely the contact,
\begin{equation}
  V_\mrm{c}(\bi{r},\bi{r}') \propto \delta^3(\bi{r}-\bi{r}'),
  \label{eq:contact_int}
\end{equation}
monopolar~\cite{ODell00a},
\begin{equation}
  V_\mrm{m}(\bi{r},\bi{r}') \propto \frac1{|\bi{r}-\bi{r}'|},
  \label{eq:monopolar_int}
\end{equation}
and dipolar interaction~\cite{Lahaye09a},
\begin{equation}
  V_\mrm{d}(\bi{r},\bi{r}') \propto
    \frac{1-3\cos^2\vartheta}{|\bi{r}-\bi{r}'|^3},
  \label{eq:dipolar_int}
\end{equation}
fulfil this requirement and therefore are possible candidates for
$\PT$-symmetric BECs.

\section{Stationary solutions}
\label{sec:spectrum}
The main goal of this paper is to gain deeper insight into the mathematical
structure of the eigenvalue spectrum of a BEC in the $\PT$-symmetric double
well~\eref{eq:potential}. We limit the following investigations to the
behaviour of the two $\PT$-symmetric and the two $\PT$-broken bound states with
the lowest real eigenvalues. This is justified because calculations of higher
excited states show a similar behaviour but at significantly higher gain/loss
contributions. In the parameter range considered the higher excited states are
only merely affected by the imaginary part of the potential.

\subsection{Eigenvalues and eigenfunctions}
\label{sec:spectrum_nonanalytic}
\begin{figure}%
  \centering%
  \includegraphics[width=0.8\columnwidth]{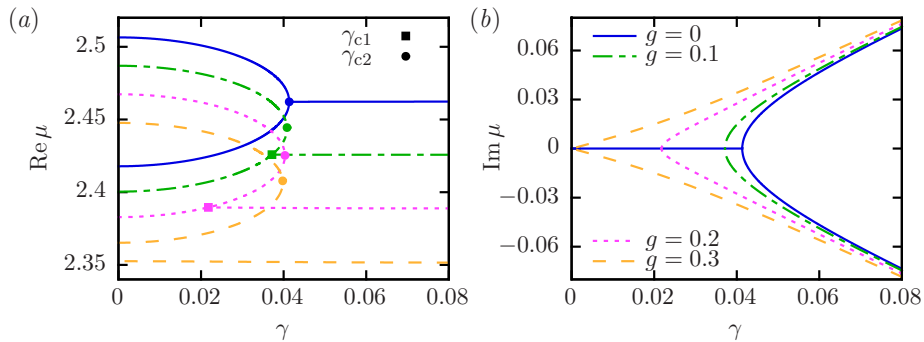}%
  \caption{Real \pt(a) and imaginary \pt(b) parts of the chemical potential
    $\mu$ for different interaction strengths $g$. In the presence of
    nonvanishing $g$ the $\PT$-symmetric solutions with real eigenvalues vanish
    at $\gamma_\mrm{c2}$ while the $\PT$-broken states with complex conjugate
    eigenvalues emerge from the ground state at
    $\gamma_\mrm{c1}<\gamma_\mrm{c2}$. Consequently a parameter region arises
    in which all states coexist.}%
  \label{fig:spectrum}%
\end{figure}

The eigenvalues and eigenfunctions have already been presented in
\cite{Dast12a}. The basic results necessary to understand the following
discussions are briefly recapitulated. The eigenvalue spectrum of the ground
state and first excited state is shown in \fref{fig:spectrum} for different
values of the nonlinearity parameter $g$.
The linear case $g=0$ shows the usual properties known from $\PT$-symmetric
systems. The two $\PT$-symmetric solutions with real eigenvalues coalesce in a
tangent bifurcation and for larger values of the gain/loss parameter $\gamma$
two $\PT$-broken solutions with complex conjugate eigenvalues arise.

For nonvanishing contact interactions we again expect a spectrum that contains
solutions with real or complex conjugate eigenvalues. \Fref{fig:spectrum}
confirms this prediction. There is, however, a crucial difference to the linear
case. The $\PT$-symmetric solutions coalesce and vanish at a critical value
$\gamma=\gamma_\mrm{c2}$ in a tangent bifurcation while the $\PT$-broken states
exist already at smaller values of $\gamma$ and emerge from the ground state at
$\gamma_\mrm{c1}$ in a pitchfork bifurcation. The physical reason for this
behaviour is the stronger localization of the $\PT$-broken states which due to
the attractive contact interaction leads to an energy reduction relative to the
$\PT$-symmetric states. It is worth noting that for a repulsive interaction,
i.e.\ $g<0$, the $\PT$-broken states are shifted upwards and emerge from the
first excited state.

The eigenvalue structure can be divided into the following regions. In the
interval $\gamma<\gamma_\mrm{c1}$ only $\PT$-symmetric solutions exist, whereas
in the regime $\gamma>\gamma_\mrm{c2}$ only the two $\PT$-broken solutions are
present. In the intermediate region the $\PT$-symmetric and $\PT$-broken states
coexist. This region grows for stronger nonlinearities $g$ and for
$g\gtrsim0.24$ the $\PT$-broken states exist even at $\gamma=0$. The existence
of symmetry breaking states at $\gamma=0$ is known as macroscopic quantum
self-trapping~\cite{Albiez05a}.

\begin{figure}%
  \centering%
  \includegraphics[width=0.8\columnwidth]{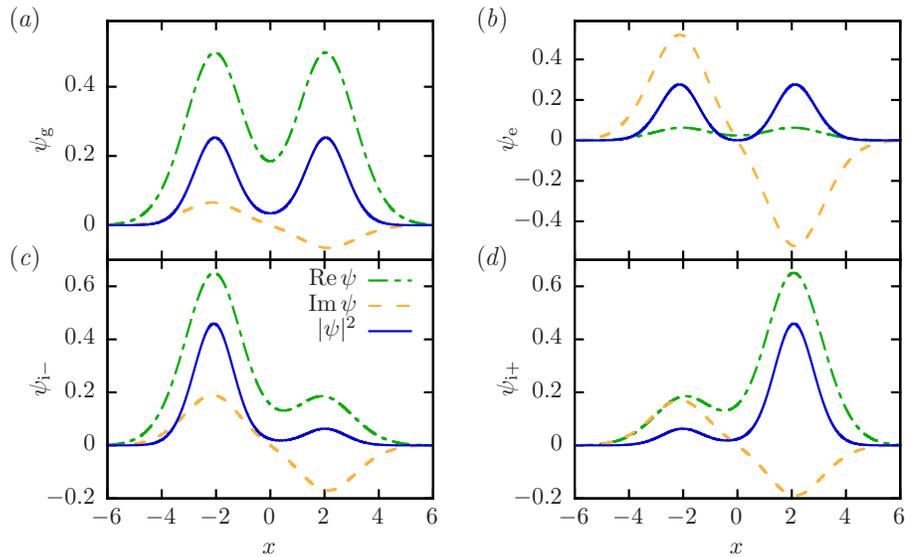}%
  \caption{The wave functions of the $\PT$-symmetric ground state \pt(a) and
    first excited state \pt(b) at $\gamma=0.01$, $g=0.2$ and the $\PT$-broken
    states with negative \pt(c) and positive imaginary part \pt(d) of the
    chemical potential at $\gamma=0.05$, $g=0.2$.}%
  \label{fig:wavefnc}%
\end{figure}

As shown in \sref{sec:pt} the eigenvalues determine the symmetries of the
corresponding wave functions. We immediately see in \fref{fig:wavefnc} that
real eigenvalues coincide
with $\PT$-symmetric wave functions. The states with complex eigenvalues have
asymmetric, and thus $\PT$-broken, wave functions. They can be mapped on each
other by application of the $\PT$ operator. The square modulus of the
$\PT$-symmetric solutions is symmetric whereas it is asymmetric for the
$\PT$-broken solutions thus breaking the symmetry of the system.

\subsection{Analytic continuation}
\label{sec:continuation}
The eigenvalue spectrum has the unusual property that the number of solutions
is not conserved when the states undergo bifurcations even though complex
solutions are also taken into account. Previous works showed that the
nonanalytic nonlinearity of the GPE, which is proportional to the square
modulus of the wave function $|\psi|^2$, is responsible for the change in the
number of solutions at the bifurcation
points~\cite{Cartarius08a,Dast12a,Gutoehrlein13a}.

It is possible to apply an analytic continuation in such a way that the number
of solutions is conserved. This allows for a better understanding of the
mathematical structure of the solutions, and in particular of the bifurcation
scenarios and exceptional points. Also the analytic continuation is used to
compare our solutions to a linear matrix model.

To apply the analytic continuation~\cite{Cartarius08a}, the complex GPE is
split into its real and imaginary parts,
\numparts
\begin{eqnarray}
  \eqlabel{eq:GPE_ce}
  &&\left[ -\Delta + V_\mrm{R}-g \left(\psi_\mrm{R}^2
  +\psi_\mrm{I}^2\right)\right]
  \psi_\mrm{R} - V_\mrm{I} \psi_\mrm{I}
  = \mu_\mrm{R}\psi_\mrm{R}-\mu_\mrm{I}\psi_\mrm{I},\\
  &&\left[ -\Delta +V_\mrm{R}-g \left(\psi_\mrm{R}^2
  +\psi_\mrm{I}^2\right)\right]
  \psi_\mrm{I} + V_\mrm{I} \psi_\mrm{R}
  = \mu_\mrm{R}\psi_\mrm{I}+\mu_\mrm{I}\psi_\mrm{R},
\end{eqnarray}
\endnumparts
with the real and imaginary parts denoted by $\mrm{R}$ and $\mrm{I}$,
respectively.

If we allow for complex values for the real and imaginary part of the wave
function and the chemical potential the two coupled differential equations are
analytic and the number of solutions is conserved. Since we have to clearly
distinguish between the usual imaginary unit $\rmi$ and the complex extension
in the calculations, a new imaginary unit $\rmj^2=-1$ is introduced to avoid
any confusion. Then the complexified real and imaginary parts of the wave
function read
\numparts
\begin{eqnarray}
  \psi_\mrm{R} &\rightarrow& \psi_1 + \rmj \psi_\rmj,
  \label{eq:bicomp_psiR}\\
  \psi_\mrm{I} &\rightarrow& \psi_\rmi + \rmj \psi_\rmk,
  \label{eq:bicomp_psiI}
\end{eqnarray}
\endnumparts
and analogously for $\mu_\mrm{R}$ and $\mu_\mrm{I}$. Actually this is identical
to extending the wave function in the original GPE to a bicomplex number with
four components,
\begin{equation}
  \psi = \psi_1 + \rmj \psi_\rmj + \rmi \psi_\rmi + \rmk \psi_\rmk,
  \label{eq:psi_bicomplex}
\end{equation}
with $\rmk=\rmi\rmj$~\cite{Gutoehrlein13a}. Bicomplex numbers are
four-dimensional hypercomplex numbers with a commutative multiplication which
allow an elegant description of the analytic continuation that complexifies
real and imaginary part.

We now have to distinguish between the complex conjugation with respect to
$\rmi$ and $\rmj$, represented by the operators $\mathcal{T}_\rmi$ and
$\mathcal{T}_\rmj$. The operator $\mathcal{T}_\rmi$ again has the physical
interpretation of time reversal and its action in coordinate space is reduced
to $\rmi\rightarrow-\rmi$. Analogously we define $\mathcal{T}_\rmj$ as the
complex conjugation $\rmj\rightarrow-\rmj$ which, however, has no inherent
physical interpretation. The system now has two symmetries, the usual
$\PT_\rmi$ symmetry and a new symmetry $\mathcal{T}_\rmj$. This leads to the
following properties. For $\PT_\rmi$-symmetric wave functions the $\mu_\rmi$
and $\mu_\rmk$ components must vanish. If one of these components has a finite
value then this value must occur with positive and negative sign and the
corresponding $\PT_\rmi$-broken wave functions can be mapped on each other by
application of the $\PT_\rmi$ operator. The same holds for the
$\mathcal{T}_\rmj$ operator for the components $\mu_\rmk$ and $\mu_\rmj$ and
additionally for the combined operator $\PT_\rmi\mathcal{T}_\rmj$ for the
components $\mu_\rmj$ and $\mu_\rmi$. This is summarized in
\tref{tab:bicomplex_sym}.
\begin{table}
  \caption{\label{tab:bicomplex_sym}
    A wave function with a specific symmetry can only have two nonvanishing
    components given by the second column. The components in the third column
    must vanish. If the wave function breaks a symmetry, the components in the
    third column occur in pairs with the same absolute value but different
    sign, and the corresponding wave functions can be mapped on each other by
    application of the respective symmetry operator.}
  \begin{indented}
  \item[]\begin{tabular}{@{}lll}
    \br
    symmetry & nonvanishing & vanishing\\
    \mr
    $\PT_\rmi$ & $\mu_1$, $\mu_\rmj$ & $\mu_\rmi $, $\mu_\rmk$\\
    $\mathcal{T}_\rmj$ & $\mu_1$, $\mu_\rmi $ & $\mu_\rmk$, $\mu_\rmj$\\
    $\PT_\rmi\mathcal{T}_\rmj$ & $\mu_1$, $\mu_\rmk$ & $\mu_\rmj$, $\mu_\rmi$\\
    \br
  \end{tabular}
  \end{indented}
\end{table}

\subsection{Analytically continued spectrum} 
\label{sec:cont_spectrum}
\begin{figure}%
  \centering%
  \includegraphics[width=0.8\columnwidth]{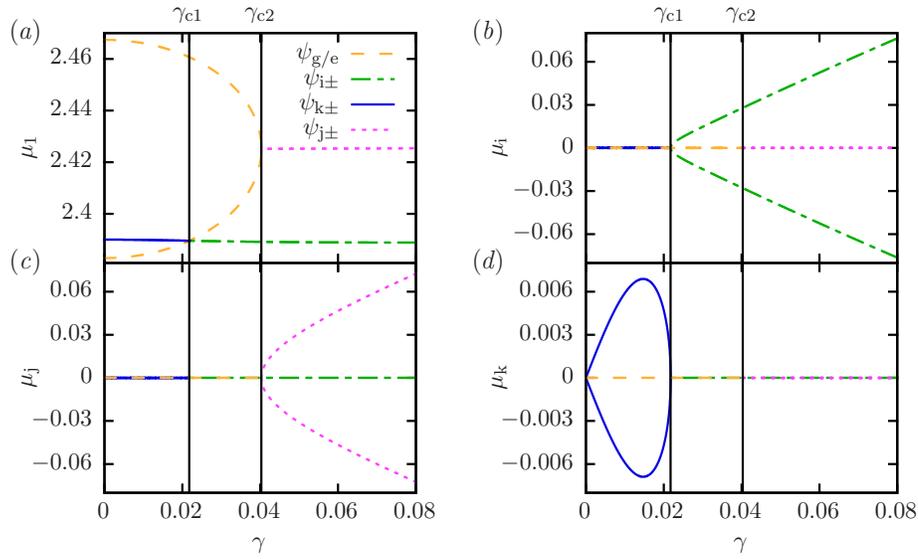}%
  \caption{The bicomplex chemical potential
    $\mu=(\mu_1+\rmj\mu_\rmj)+\rmi(\mu_\rmi+\rmj\mu_\rmk)$ of the eigenstates
    with the analytic continuation at a nonlinearity $g=0.2$. The solutions
    $\psi_\mrm{g/e}$ with real and $\psi_{\rmi\pm}$ with complex eigenvalues
    are not changed by the continuation. Two additional branches
    $\psi_{\rmk\pm}$ ($\psi_{\rmj\pm}$) arise below $\gamma_\mrm{c1}$ (above
    $\gamma_\mrm{c2}$), therefore the number of solutions is conserved for all
  values of $\gamma$.}%
  \label{fig:bicomplex_spectrum}%
\end{figure}

Using bicomplex wave functions the chemical potential $\mu$ of the eigenstates
also becomes a bicomplex number with four components
\begin{equation}
  \mu=(\mu_1+\rmj\mu_\rmj)+\rmi(\mu_\rmi+\rmj\mu_\rmk).
  \label{eq:bicomplex_mu}
\end{equation}
\Fref{fig:bicomplex_spectrum} shows that the GPE now has four solutions
independent of the parameter $\gamma$ and thus the number of solutions is
constant, as expected. The already known solutions occur unchanged in the
spectrum.

For the $\PT_\rmi$-symmetric states with real eigenvalues existing for
$\gamma<\gamma_\mrm{c2}$ the value of $\mu_1$ is identical to the real part of
$\mu$ without the analytic continuation. The other three components vanish
$\mu_\rmj=\mu_\rmi=\mu_\rmk=0$. These solutions coalesce as before in an
tangent bifurcation at $\gamma_\mrm{c2}$. However, they do not vanish at this
point. They are continued to the regime $\gamma>\gamma_\mrm{c2}$ by two states
that only exist within the complex extension. These two additional states have
nonzero $\mu_1$ and $\mu_\rmj$ components, i.e.\ their chemical potentials have
complexified real parts $\mu=\mu_1+\rmj\mu_\rmj$. 

At $\gamma_\mrm{c1}$ the branches with the already known $\PT_\rmi$-broken
solutions emerge from the ground state in a pitchfork bifurcation. The chemical
potential of these states has the form $\mu=\mu_1+\rmi\mu_\rmi$. On these
branches the chemical potential is a usual complex number because the
components $\mu_\rmj$ and $\mu_\rmk$ are zero. Since the number of solutions is
also not changed at $\gamma_\mrm{c1}$, two additional branches arise in the
interval $0<\gamma<\gamma_\mrm{c1}$ which continue the $\PT_\rmi$-broken
solutions to values of $\gamma$ smaller than $\gamma_\mrm{c1}$. The eigenvalues
on these branches have the form $\mu=\mu_1+\rmk\mu_\rmk$.

\begin{figure}%
  \centering%
  \includegraphics[width=0.8\columnwidth]{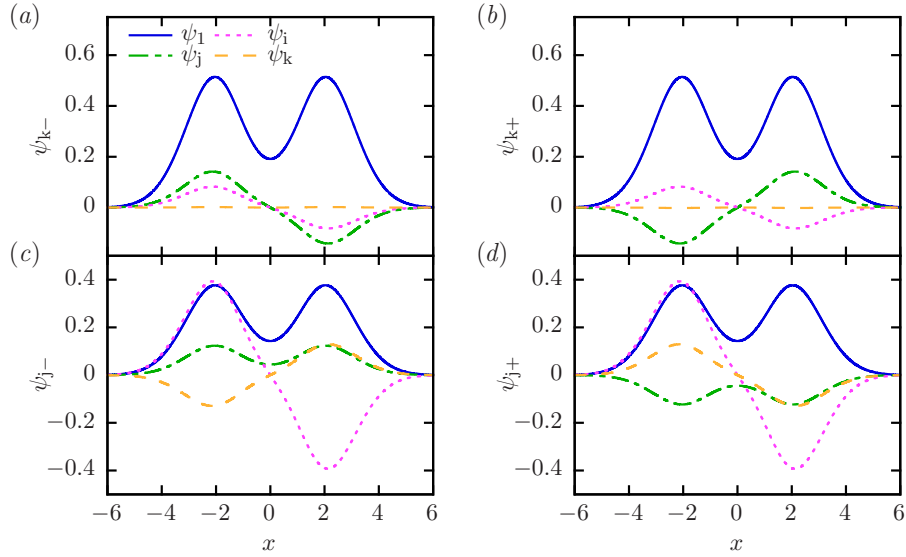}%
  \caption{The bicomplex wave functions on the continued branches at a
    nonlinearity $g=0.2$ and a gain/loss parameter $\gamma=0.01$ \pt(a),\pt(b)
    and $\gamma=0.05$ \pt(c),\pt(d), respectively. The wave functions on the
    branches $\psi_{\rmk\pm}$, which arise at the pitchfork bifurcation at
    $\gamma_\mrm{c1}$, are $\PT_\rmi \mathcal{T}_\rmj$-symmetric but $\PT_\rmi$
    and $\mathcal{T}_\rmj$-broken \pt(a),\pt(b). On the second continued
    branches $\psi_{\rmj\pm}$, which emerge at $\gamma_\mrm{c2}$, the wave
    functions are $\PT_\rmi$-symmetric but break the $\mathcal{T}_\rmj$ and
    $\PT_\rmi\mathcal{T}_\rmj$ symmetry \pt(c),\pt(d).}%
  \label{fig:bicomplex_wavefnc}%
\end{figure}

The wave functions on the new branches $\psi_{\rmk\pm}$ and $\psi_{\rmj\pm}$
are shown in \fref{fig:bicomplex_wavefnc}. 
To discuss the symmetries of the eigenstates one should point out the actions
of the symmetry operators on the wave functions' components. The $\PT_\rmi$
operator reflects all components at $x=0$ and additionally the $\psi_\rmi$ and
$\psi_\rmk$ components at the $x$ axis. The $\mathcal{T}_\rmj$ operator only
reflects the $\psi_\rmj$ and $\psi_\rmk$ components at the $x$ axis. The
combined action is represented by the $\PT_\rmi\mathcal{T}_\rmj$ operator which
reflects all components at $x=0$ and the $\psi_\rmi$ and $\psi_\rmj$ components
at the $x$ axis.

The branches $\psi_{\rmk\pm}$ with nonzero $\mu_1$ and $\mu_\rmk$ components
have, according to \tref{tab:bicomplex_sym}, wave functions with $\PT_\rmi
\mathcal{T}_\rmj$ symmetry and broken $\PT_\rmi$ and $\mathcal{T}_\rmj$
symmetry. This can be verified by looking at the corresponding wave functions
in figures~\ref{fig:bicomplex_wavefnc}\pt(a) and
\ref{fig:bicomplex_wavefnc}\pt(b). The two wave functions are mapped on
themselves by application of the $\PT_\rmi\mathcal{T}_\rmj$ and are mapped on
each other by application of the $\PT_\rmi$ or $\mathcal{T}_\rmj$ operator.
Therefore we can conclude that the branches with complex eigenvalues are
continued to the interval $\gamma<\gamma_\mrm{c1}$ using the analytic
continuation, and the property is conserved that they are $\PT_\rmi$-broken.
However, there is a significant difference. The $\PT_\rmi$-broken branches
$\psi_{\rmi\pm}$ in the interval $\gamma>\gamma_\mrm{c1}$ have asymmetric
eigenstates, whereas the wave functions of the continued branches have
symmetric and antisymmetric components although the complete wave functions are
$\PT_\rmi$-broken. The square modulus appearing in the GPE
\begin{equation}
  \left|\psi\right|^2=\psi_\mrm{R}^2+\psi_\mrm{I}^2
  =(\psi_1+\rmj\psi_\rmj)^2 + (\psi_\rmi+\rmj\psi_\rmk)^2
  \label{eq:bicomplex_squaremod}
\end{equation}
has a symmetric $1$-component and an antisymmetric $\rmj$-component for the
branches $\psi_{\rmk\pm}$. Thus the square modulus is asymmetric, which is
consistent with the wave functions being $\PT_\rmi$-broken.

The figures~\ref{fig:bicomplex_wavefnc}\pt(c) and
\ref{fig:bicomplex_wavefnc}\pt(d) show the wave functions on the branches which
arise at the pitchfork bifurcation at $\gamma_\mrm{c2}$. Because only the
$\mu_1$ and $\mu_\rmj$ components are nonzero, comparison with
\tref{tab:bicomplex_sym} shows that the eigenstates must obey $\PT_\rmi$
symmetry, but are $\mathcal{T}_\rmj$-broken and hence also
$\PT_\rmi\mathcal{T}_\rmj$-broken. Therefore the wave functions are invariant
under the action of the $\PT_\rmi$ operator and are mapped on each other by the
$\mathcal{T}_\rmj$ and $\PT_\rmi\mathcal{T}_\rmj$ operator. This means
$\PT_\rmi$-symmetric solutions exist for all values of $\gamma$ if the analytic
continuation is applied. For all $\PT_\rmi$-symmetric solutions the square
modulus is symmetric, which can be easily checked in
\eref{eq:bicomplex_squaremod}.

The properties of the additional symmetries can also be applied to the already
known $\PT_\rmi$-symmetric and broken states but this only leads to trivial
results because the $\rmj$ and $\rmk$ components of these states are always
zero.

It must be emphasized that although the analytic continuation allows us to gain
a deeper insight into the mathematical structure, the additional states have no
physical meaning. In the following the $\mathcal{T}_\rmj$ symmetry is not
further discussed and we again write $\PT$ instead of $\PT_\rmi$.

\section{Exceptional points}
\label{sec:ep}
At an exceptional point two or more eigenvalues and the corresponding
eigenvectors of a non-Hermitian quantum system are degenerate. An exceptional
point at which $n$ eigenvalues and eigenvectors coalesce is called an $n$-th
order exceptional point (EP$n$). The characteristic signature of an EP$n$ is
the cyclic permutation of the $n$ eigenvalues and eigenvectors while encircling
the exceptional point in the complex parameter space.

Two exceptional points are found in the eigenvalue structure of a BEC in a
$\PT$-symmetric double-delta-trap~\cite{Cartarius12a,Heiss13a} which shows a
similar behaviour as the system studied in this paper. Using the
delta-potential the point at $\gamma_\mrm{c2}$ at which the ground state and
the excited state coalesce was found to be a second-order exceptional point
(EP2). A second exceptional point arises due to the contact interaction at
$\gamma_\mrm{c1}$ where the $\PT$-broken states emerge from the ground state.
Here three eigenvalues and eigenstates coalesce and the exceptional point has
the characteristic behaviour of an EP3. In this section we show that the same
exceptional points can also be found in the spatially extended double-well
potential. However, in contrast to the above mentioned results here we
distinguish between the four components of the continuation and do not use a
simplified approach with a projection to two components. 

Before treating the exceptional points of the full problem with complex
potential and nonlinearity the linear limit ($g = 0$) is investigated.

\subsection{Exceptional point in the linear case}
\label{sec:ep_linear}
In the linear case the spectrum has only one exceptional point $\gamma_\mrm{c}$
at which the ground and excited state coalesce and complex solutions arise.
The characteristic properties of exceptional points are studied by encircling
the exceptional point with a complex parameter.

In the case of the critical point at $\gamma_\mrm{c}$ we expand the real
parameter $\gamma$ to a complex number. One cycle with radius $r$ is
parameterized by
\begin{equation}
  \gamma=\gamma_\mrm{c} + r\rme^{\rmi\vartheta},\qquad \vartheta=0,\dots,2\pi.
  \label{eq:ep2_lin_path}
\end{equation}
The imaginary part of the perturbation $\imag\gamma=r\sin\vartheta$ must be
interpreted as a real contribution to the potential of the system.
Inserting~\eref{eq:ep2_lin_path} into the potential~\eref{eq:potential} shows
that an additional real term
\begin{equation}
  V_\mrm{p}(x)=-\imag\gamma\, x\rme^{-\rho x^2}
  \label{eq:asymmetric_pot}
\end{equation}
occurs. Thus the imaginary part of $\gamma$ breaks the $\PT$ symmetry of the
potential, because the symmetric real part of the potential obtains an
antisymmetric contribution proportional to $\imag\gamma$.

\begin{figure}%
  \centering%
  \includegraphics[width=0.8\columnwidth]{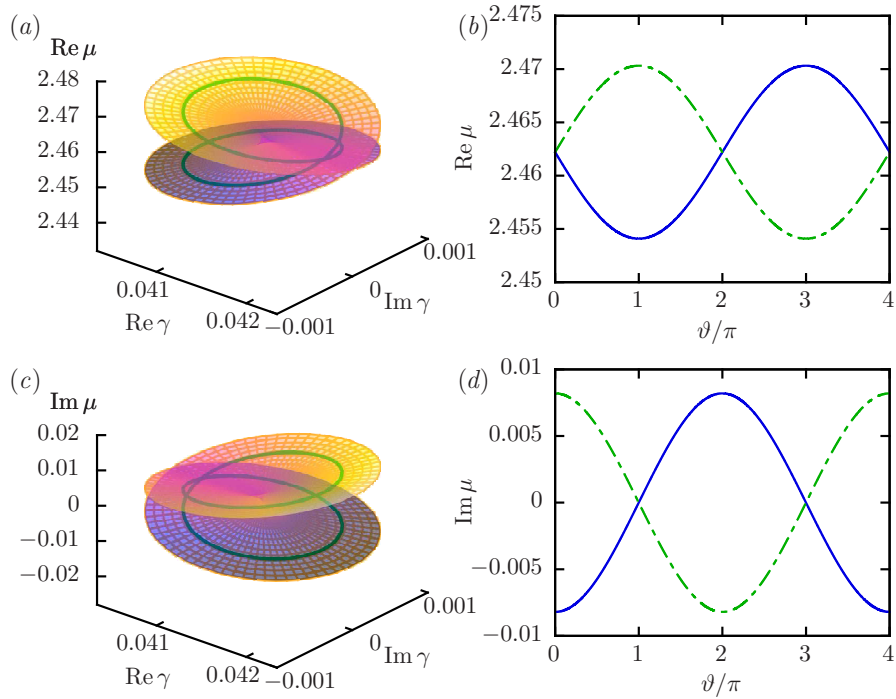}%
  \caption{Real \pt(a),\pt(b) and imaginary parts \pt(c),\pt(d) of the
    eigenvalues in the vicinity of the exceptional point at $\gamma_\mrm{c}$ in
    the linear case $g=0$. The cycle highlighted in the left figures is
    parameterized by \eref{eq:ep2_lin_path} with $r=7\times10^{-4}$. The
    behaviour of the eigenvalues on this path as a function of $\vartheta$ is
    shown in \pt(b) and \pt(d), where different line types distinguish the two
    eigenvalues. The exceptional point shows the characteristic behaviour of an
    EP2.}%
  \label{fig:ep2_lin}%
\end{figure}

The behaviour of the energy eigenvalues in the vicinity of the exceptional
point is shown in \fref{fig:ep2_lin}\pt(a),\pt(c). The two eigenvalues involved
in the
exceptional point are represented by two surfaces (Riemann sheets). Following a
path around the exceptional point leads to a transition from the upper to the
lower surface and vice versa. The cycle with fixed radius $r=7\times10^{-4}$
shown in \fref{fig:ep2_lin}\pt(b),\pt(d) starts on the branches with
$\PT$-broken solutions with complex eigenvalues at $\vartheta=0$. Here the
eigenvalues have the same real part but a positive or negative imaginary part.
At $\vartheta=\pi$ the parameter $\gamma$ is real again, but on the branches
with $\PT$-symmetric solutions on the other side of the exceptional point.
After one cycle each state has turned into the respective other state, and
after two cycles the initial configuration is restored. This is consistent with
the square root behaviour of an EP2.

\subsection{Exceptional points in the nonlinear case}
\label{sec:ep_nonlinear}
\begin{figure}%
  \centering%
  \includegraphics[width=0.8\columnwidth]{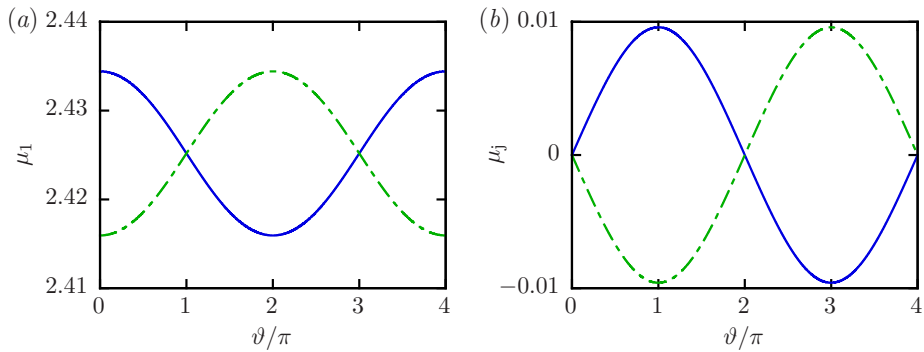}%
  \caption{The components $\mu_1$ \pt(a) and $\mu_\rmj$ \pt(b) of the two
    eigenvalues, distinguished by different line types, while encircling the
    critical point at $\gamma_\mrm{c2}$ on the path parameterized in
    \eref{eq:ep2_nonlin_path} with the nonlinearity $g=0.2$ and the radius
    $r=0.001$. The two additional components $\mu_\rmi$ and $\mu_\rmk$ are zero
    on the whole cycle. The eigenvalues show the characteristic
    square root behaviour of an EP2.}%
  \label{fig:ep2_nonlin}%
\end{figure}

Encircling the exceptional points is apparently only possible if the number of
states is conserved and therefore in the nonlinear case the analytic
continuation has to be used. The eigenvalue spectrum shows two exceptional
points at $\gamma_\mrm{c1}$ and $\gamma_\mrm{c2}$. We first discuss the
exceptional point at $\gamma_\mrm{c2}$ where the ground and excited state
coalesce. One cycle around the exceptional point is parameterized with a
complexified real part of the gain/loss parameter,
\begin{equation}
  \gamma=\gamma_\mrm{c2}-r\rme^{\,\rmj\vartheta},
  \qquad \vartheta=0,\dots,2\pi.
  \label{eq:ep2_nonlin_path}
\end{equation}
The path starts at $\gamma=\gamma_\mrm{c2}-r$ in the regime of entirely real
eigenvalues. After half a cycle the parameter $\gamma=\gamma_\mrm{c2}+r$ is
real again, but on the other side of the exceptional point. At this value the
two additional states with nonzero $\mu_1$ and $\mu_\rmj$ components exist. On
the whole cycle the two components $\mu_\rmi$ and $\mu_\rmk$ are zero. As can
be seen in \fref{fig:ep2_nonlin}
the two eigenvalues and respective eigenvectors interchange after one cycle,
and after two cycles the initial configuration is restored. This is the
characteristic square root behaviour of an EP2.

At the exceptional point at $\gamma_\mrm{c1}$ three eigenvalues and
eigenvectors become equal. This exceptional point is encircled on two different
paths. On the one hand a complexified real part of the nonlinearity parameter
is used,
\begin{equation}
  g = g_\mrm{c1} + r\rme^{\,\rmj\vartheta},
  \qquad \vartheta=0,\dots,2\pi,
  \label{eq:ep32_nonlin_path}
\end{equation}
where the exceptional point resides at the real value $g_\mrm{c1}$. On the
other hand a complexified imaginary part of the gain/loss parameter is used,
\begin{equation}
  \gamma=\gamma_\mrm{c1}+\rmi r\rme^{\,\rmj\vartheta}
  \equiv \gamma_\mrm{c1}+\rmi\varepsilon,
  \qquad \vartheta=0,\dots,2\pi,
  \label{eq:ep33_nonlin_path}
\end{equation}
which again breaks the symmetry of the real part of the potential. The
symmetry-breaking contribution is controlled by the asymmetry parameter
$\varepsilon$.

\begin{figure}%
  \centering%
  \includegraphics[width=0.8\columnwidth]{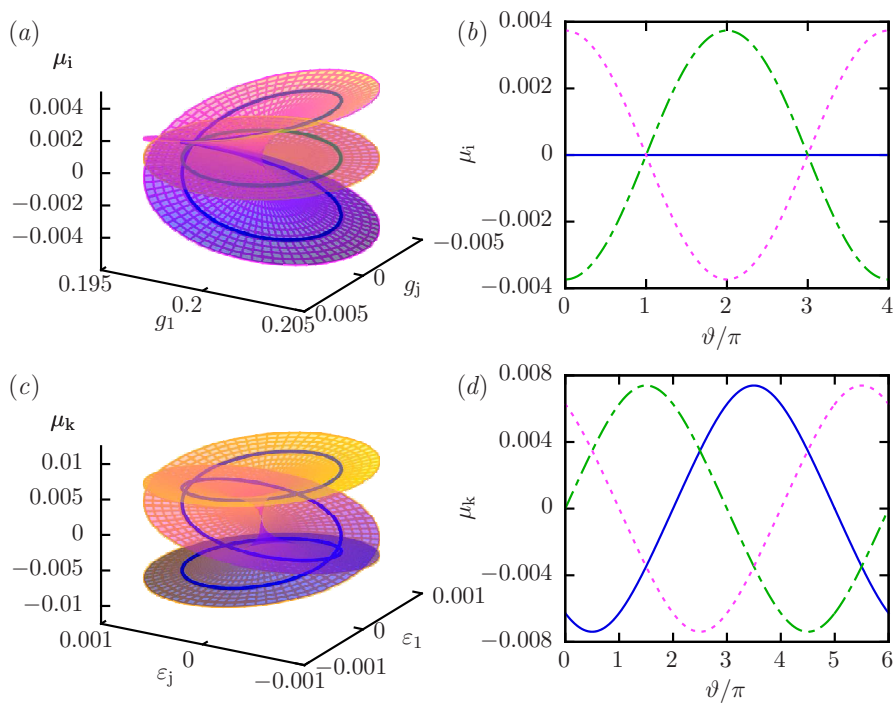}%
  \caption{The behaviour of the energy eigenvalues while encircling the
    exceptional point at $\gamma_\mrm{c1}$ and $g=0.2$. The perturbed
    nonlinearity defined in \eref{eq:ep32_nonlin_path} results in the
    characteristic square root behaviour of an EP2 \pt(a),\pt(b). Cyclic
    permutations of the three eigenvalues \pt(c),\pt(d) are gained with the
    complex expanded gain/loss parameter \eref{eq:ep33_nonlin_path}. This
    confirms that the exceptional point at $\gamma_\mrm{c1}$ is indeed an EP3.
    The right figures show the eigenvalue component as a function of
    $\vartheta$ on the path highlighted in the left figures. Again different
    line types denote the three different eigenvalues. For the sake of brevity
    only one of the four $\mu$ components is shown for each path, but the other
    three components show an analogous behaviour.}%
  \label{fig:ep3_nonlin}%
\end{figure}

The evolution of the eigenvalues while encircling the exceptional point with
the perturbed nonlinearity \eref{eq:ep32_nonlin_path} is shown in the
figures~\ref{fig:ep3_nonlin}\pt(a) and \ref{fig:ep3_nonlin}\pt(b).
Figure~\ref{fig:ep3_nonlin}\pt(a) shows the component $\mu_\rmi$ of the
eigenvalues in the vicinity of the exceptional point. Each surface corresponds
to one of the three eigenvalues. For real values of the parameter $g$ the
surfaces can be identified with states of the known eigenvalue spectrum. The
middle surface represents the $\PT$-symmetric ground state $\psi_\mrm{g}$ with
real eigenvalues. The top (bottom) surface corresponds to the $\PT$-broken
state $\psi_{\rmi+}$ ($\psi_{\rmi-}$) with positive (negative) imaginary part.
One recognizes that a path starting on the ground state does not leave its
surface, i.e.\ the ground state does not interchange with another state while
encircling the exceptional point. The behaviour of the $\PT$-broken states
corresponds to that of an EP2. After one cycle the states are interchanged, and
after two cycles the initial configuration is restored. It is known that for
certain perturbations an EP3 can show such an EP2 signature~\cite{Graefe12a}.

To confirm that this point is indeed an EP3 it is necessary to find a
perturbation for which the three eigenvalues and eigenvectors interchange
cyclically. This behaviour is observed on the path parameterized by
\eref{eq:ep33_nonlin_path}. Figures~\ref{fig:ep3_nonlin}\pt(c) and
\ref{fig:ep3_nonlin}\pt(d) show the eigenvalue component $\mu_\rmk$ in the
proximity of the exceptional point for the whole parameterized region \pt(c)
and for a cycle with specific radius \pt(d). Starting on the bottom Riemann
sheet, after one counterclockwise cycle the path lies on the middle sheet,
after two circles on the top sheet, and finally after three circles on the
bottom sheet again. This corresponds to the expected cubic root behaviour of an
EP3.

\section{Linear matrix model}
\label{sec:matrixmodel}
The results presented so far have been obtained with the {\em nonlinear}
GPE \eref{eq:gpe}.
We now want to show that quantitatively very similar results can be
obtained from a simple {\em linear} matrix model.
Our starting point is a nonlinear model system introduced by Graefe
\etal \cite{Graefe10,Graefe12b}, which shows a very good qualitative
agreement with the properties of our system, viz.\ a two mode
approximation of a BEC in a double well.
The chemical potential of this approximation of the BEC is given as
\begin{equation}
 \tilde{\mu} = 2(\tilde{\varepsilon}
 -\rmi\tilde{\gamma})s_z + 4\tilde{g} s_z^2 + 2v s_x,
\label{eq:mu_model}
\end{equation}
with the vector $\bi{s}=(s_x,s_y,s_z)$ the stationary solution 
(fixed point) of the nonlinear and non-Hermitian Bloch equations
\numparts
\begin{eqnarray}
  \eqlabel{eq:bloch}
  \dot s_x &=& -2\tilde{\varepsilon} s_y - 4\tilde{g} s_y s_z 
  + 4\tilde{\gamma} s_x s_z, \\
  \dot s_y &=& 2\tilde{\varepsilon} s_x + 4\tilde{g} s_x s_z -2v s_z 
  + 4\tilde{\gamma} s_y s_z, \\
  \dot s_z &=& 2v s_y - \tilde{\gamma} (1 - 4s_z^2).
\end{eqnarray}
\endnumparts
The parameter $v$ defines the coupling between the two wells, and
$\tilde\gamma$ corresponds to the gain/loss parameter $\gamma$ of the spatially
extended potential. The strength of the nonlinearity $g$ and the symmetry
breaking potential $\varepsilon$ are represented by $\tilde g$ and
$\tilde\varepsilon$ in the model. We have derived a linear matrix model, whose
eigenvalues exactly agree with the solution of the nonlinear model. The
$4\times4$ matrix reads
\begin{equation}
 H = \left(\begin{array}{cccc}
 0 & 1 & 0 & 0 \\
 v^2-\tilde{\gamma}^2+\tilde{\varepsilon}^2 & 0 & 1 & 0 \\
 -\frac{\rmi \tilde{g}^3 \tilde{\gamma}\tilde{\varepsilon}}
 {\tilde{g}^2+\tilde{\gamma}^2} &
 -4\rmi \tilde{\gamma}\tilde{\varepsilon}
 + \frac{4\tilde{\gamma}^2-2\tilde{g}^2}{\tilde{g}^2
 +\tilde{\gamma}^2}\tilde{\varepsilon}^2 &
 \tilde{g}+\frac{2\rmi \tilde{g} \tilde{\gamma}\tilde{\varepsilon}}
 {\tilde{g}^2+\tilde{\gamma}^2} & 1 \\
 h_{41} & -\frac{\rmi \tilde{g}^3 \tilde{\gamma}\tilde{\varepsilon}}
 {\tilde{g}^2+\tilde{\gamma}^2} & h_{43} & \tilde{g}
 \end{array} \right)
\label{eq:H_model}
\end{equation}
with the matrix elements
\begin{eqnarray*}
 h_{41} &=& \frac{-\rmi \tilde{\gamma}\tilde{\varepsilon}}
 {\tilde{g}^2+\tilde{\gamma}^2}
 \left[\tilde{g}^4+(2\tilde{g}^2+4\tilde{\gamma}^2)
 (\tilde{\gamma}^2-v^2)\right]\\
 &&+\frac{2\tilde{\gamma}^2-\tilde{g}^2}{\tilde{g}^2+\tilde{\gamma}^2}
 \left[(4\tilde{\gamma}^2+\tilde{g}^2-2v^2)\tilde{\varepsilon}^2
 + 2\rmi \tilde{\gamma}\tilde{\varepsilon}^3 -
 2\tilde{\varepsilon}^4\right],\\
 h_{43} &=& \frac{v^2\tilde{\gamma}^2}{\tilde{g}^2+\tilde{\gamma}^2}
 -\tilde{\gamma}^2
 + \frac{2\tilde{g}^2-3\tilde{\gamma}^2}{\tilde{g}^2+\tilde{\gamma}^2}
 \tilde{\varepsilon}^2.
\end{eqnarray*}
Details of the derivation are given in the appendix.

The eigenvalues of the linear matrix model~\eref{eq:H_model} are related not
only to the $\PT$-symmetric real and the symmetry breaking complex solutions of
the GPE \eref{eq:gpe} but also to states obtained with the analytically
continued equations \eref{eq:GPE_ce}. However, in this section we do not
compare individual bicomplex components of the chemical potential but use
the projections 
\numparts
\begin{eqnarray}
  \eqlabel{eq:mu_merge}
  \mu_+ &= (\mu_1 + \mu_\rmk) + \rmj\,(\mu_\rmj - \mu_\rmi),\\
  \mu_- &= (\mu_1 - \mu_\rmk) + \rmj\,(\mu_\rmj + \mu_\rmi)
\end{eqnarray}
\endnumparts
of the bicomplex solutions on complex numbers, which are formally
obtained by the replacement $\rmi\to\mp\rmj$ in \eref{eq:bicomplex_mu}.
For vanishing asymmetry parameter $\tilde{\varepsilon}=0$ the
eigenvalues $\tilde{\mu}_+$ and $\tilde{\mu}_-$ coincide, and
the matrix \eref{eq:H_model} simplifies to
\begin{equation}
 H = \left(\begin{array}{cccc}
 0 & 1 & 0 & 0 \\
 v^2-\tilde{\gamma}^2 & 0 & 1 & 0 \\
 0 & 0 & \tilde{g} & 1 \\
 0 & 0 & \frac{v^2\tilde{\gamma}^2}{\tilde{g}^2+\tilde{\gamma}^2}
 -\tilde{\gamma}^2 & \tilde{g} \end{array} \right),
\label{eq:H_model_eps0}
\end{equation}
with the eigenvalues
\begin{equation}
  \tilde{\mu} =
  \cases{
    \pm\sqrt{v^2-\tilde{\gamma}^2},\\
    \tilde{g}\pm\tilde{\gamma}\sqrt{\frac{v^2}{\tilde{g}^2
    +\tilde{\gamma}^2}-1}.}
\label{eq:eigenval_matrix}
\end{equation}

To compare the matrix model \eref{eq:H_model} with the numerically exact
solutions of the GPE \eref{eq:gpe} the parameters $\tilde{g}$,
$\tilde{\gamma}$, and $\tilde{\varepsilon}$ of the model are replaced with
appropriately scaled quantities, $\tilde{g}=g/g_0$,
$\tilde{\gamma}=\gamma/\gamma_0$, $\tilde{\varepsilon}=\varepsilon/\gamma_0$,
and a shift $\tilde{\mu}=\mu-\mu_0$ of the chemical potential is introduced.
\begin{figure}%
  \centering%
  \includegraphics[width=0.8\columnwidth]{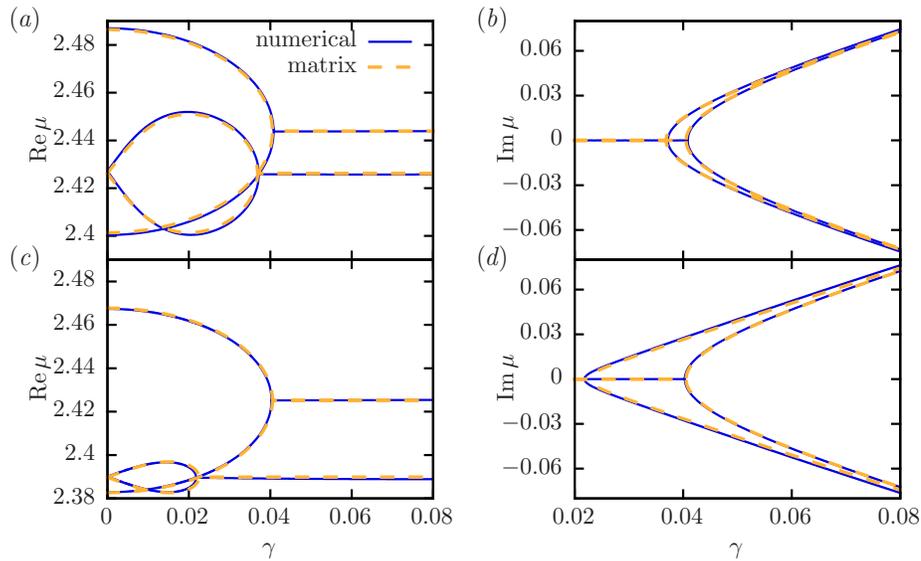}%
  \caption{Comparison of the eigenvalues \eref{eq:eigenval_matrix} of the
    matrix model using the adjusted parameters (see text) with the numerical
    results of the spatially extended double-well potential for $g=0.1$
    \pt(a),\pt(b) and $g=0.2$ \pt(c),\pt(d). The eigenvalues show an excellent
    agreement even though a wide parameter range is compared.}%
\label{fig:matrix_fit}%
\end{figure}%
\Fref{fig:matrix_fit} compares the two models for the nonlinearities $g=0.1$
and $0.2$. The values of $g_0=-5.64$, $\gamma_0= 0.953$ and $v=0.0426$,
obtained by adjusting the eigenvalues~\eref{eq:eigenval_matrix} of the matrix
model to the exact results, are independent of the nonlinearity, however, the
energy shift $\mu_0$ differs. The eigenvalues are in excellent agreement and
thus the matrix model predicts the correct eigenvalues for a wide range of
gain/loss and nonlinearity parameters.

\begin{figure}%
  \centering%
  \includegraphics[width=0.8\columnwidth]{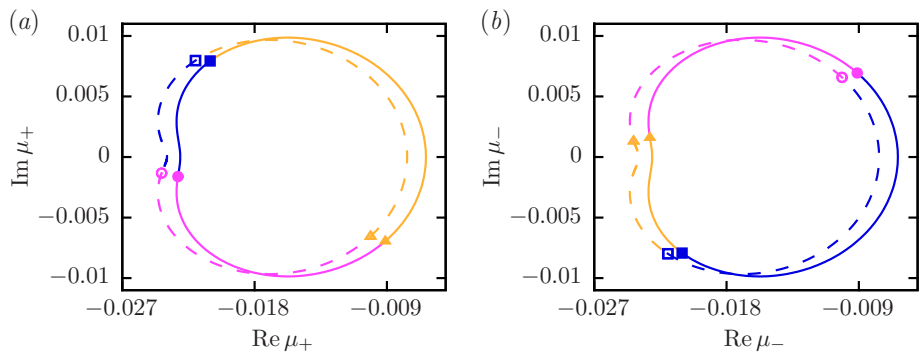}%
  \caption{Shifted eigenvalues $\mu_+$ \pt(a) and $\mu_-$ \pt(b) of the complex
    projections defined in \eref{eq:mu_merge} when encircling the exceptional
    point at $\gamma_\mrm{c1}$ along the path given in
    \eref{eq:ep33_nonlin_path}. Solid lines: Numerically exact results of the
    GPE. Dashed lines: Results of the matrix model. The three states permute
    after one cycle in the parameter space.}%
\label{fig:EP3}%
\end{figure}

The matrix model~\eref{eq:H_model} also provides the correct properties of the
exceptional points. The eigenvalues $\mu_\pm$ while encircling the exceptional
point at $\gamma_\mrm{c1}$ along the path given in \eref{eq:ep33_nonlin_path}
are shown in figures~\ref{fig:EP3}\pt(a) and \pt(b) for the nonlinearity
$g=0.2$.
The eigenvalues are shifted by the mean value of the chemical potential
obtained with the four states (including the isolated state not shown in
\fref{fig:EP3}). The solid and dashed lines present the numerically exact
results and the eigenvalues of the matrix model, respectively. Both figures
clearly exhibit the permutation of the three states after one cycle in the
parameter space indicating the EP3. Note that in the matrix
model~\eref{eq:H_model} the exceptional point at $\gamma_\mrm{c1}$ is an EP3,
whereas in the matrix model in \cite{Graefe12b} an EP2 is found, i.e.\ only two
instead of three eigenvectors coincide. Heiss \etal~\cite{Heiss13a}
constructed a three-dimensional matrix model with interacting levels that shows
an EP3 at $\gamma_\mrm{c1}$ and has three of the eigenvalues given by
\eref{eq:eigenval_matrix}. The four-dimensional matrix model~\eref{eq:H_model}
has the advantage that both the EP3 at the pitchfork bifurcation and the EP2 at
the tangent bifurcation can be described, and the model even allows us to study
interactions between all four eigenmodes in the next section.

\section{Four interacting eigenmodes}
\label{sec:EP4}
The degeneracy of two states in an EP2 can be achieved in systems which are
controlled by at least two external parameters, and has been observed in a
quite large variety of systems (see \cite{Heiss12a} and references therein).
The degeneracy of more than
two eigenvalues and eigenvectors is possible, in principle, however, an EP$n$
in general requires the adjustment of $(n^2+n-2)/2$ parameters \cite{Heiss08}.
Therefore, the occurrence of higher order exceptional points appears to be more
rare. EP3s have been observed in dipolar condensates \cite{Gutoehrlein13a} and
in the $\PT$-symmetric condensate discussed here (see sections~\ref{sec:ep} and
\ref{sec:matrixmodel}). Signatures of three interacting eigenmodes related to
an EP3 have also been found, e.g.\ in resonance spectra of the hydrogen atom in
crossed magnetic and electric fields \cite{Cartarius09} and two dielectric
microdisks \cite{Ryu12}. Optical wave guides can be arranged such that they
contain EP3s~\cite{Graefe12a}. Higher order exceptional points occur in a
non-Hermitian $\PT$-symmetric Bose-Hubbard model~\cite{Graefe08a}.
However, to the best of our knowledge more than three interacting eigenmodes
have not yet been observed in any non-Hermitian quantum system describing a
single particle or a BEC with a mean-field approach.

The pitchfork bifurcation at $\gamma_\mrm{c1}$ and the tangent bifurcation at
$\gamma_\mrm{c2}$ merge in the limit $g\to 0$. At $\tilde{g}=0$ and
$\tilde{\gamma}=v$ the four eigenvalues in \eref{eq:eigenval_matrix} are
degenerate. The matrix model~\eref{eq:H_model_eps0} at those parameters has
the form of a Jordan block of an EP4, viz.
\begin{equation}
 H = J[H] = \left(\begin{array}{cccc}
 0 & 1 & 0 & 0 \\
 0 & 0 & 1 & 0 \\
 0 & 0 & 0 & 1 \\
 0 & 0 & 0 & 0 \end{array} \right),
\label{eq:Jordan_H}
\end{equation}
which means there is only one eigenvector belonging to the four degenerate
eigenvalues $\tilde{\mu}=0$. The structure of the Jordan block in
\eref{eq:Jordan_H} strongly motivates the search for signatures of an EP4, such
as the permutation of four states after one cycle on an appropriately chosen
parameter path, in both the matrix model and in the system described by the
GPE~\eref{eq:gpe}.

To reveal the EP4 we have to take into account that the limit $g\to0$ in the
GPE is nontrivial~\cite{Heiss13a}. For $\tilde{g}=0$ the matrix $H$ in
\eref{eq:H_model} has two doubly degenerate branches of eigenvalues
$\tilde{\mu}=\pm\sqrt{v^2-(\tilde{\gamma} +\rmi\tilde{\varepsilon})^2}$. We
choose a small but nonzero nonlinearity parameter $\tilde{g}$, which lifts
that degeneracy but also causes the branching singularity to split into the
tangent and the pitchfork bifurcation. To search for signatures of the EP4 we
encircle both branching points simultaneously on paths with sufficiently large
radii.

For a path in the complex $\tilde{\gamma}$ plane (with centre
$\tilde{\gamma}=v$ and constant $\tilde{\varepsilon}=0$) and for a path in the
complex $\tilde{\varepsilon}$ plane (with centre $\tilde{\varepsilon}=0$ and
constant $\tilde{\gamma}=v$) the eigenvalues of the matrix $H$ show
qualitatively the same behaviour: The eigenvalues move in pairs on two paths
where two states permute after one cycle in the parameter space, i.e.\ they
show the typical behaviour of two EP2s related to two tangent bifurcations.

\begin{figure}%
  \centering%
  \includegraphics[width=0.8\columnwidth]{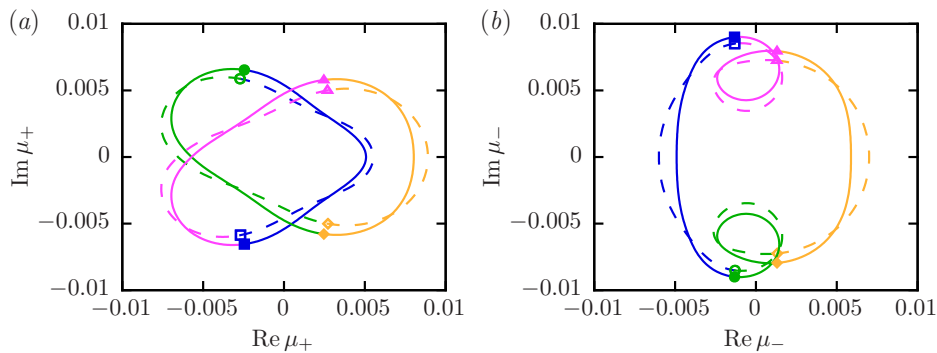}%
  \caption{Shifted eigenvalues \pt(a) $\mu_+$ and \pt(b) $\mu_-$ when
    simultaneously encircling the two exceptional points with
    parameters $g=0.02$, $\varepsilon=0.0002\rmj$ along the path
    $\gamma=\gamma_0v+0.0004\rme^{\,\rmj\vartheta}$. Solid lines: Numerically
    exact results of the GPE. Dashed lines: Results of the matrix model.
    All four states permute after one cycle in the parameter space indicating
    the existence of an EP4.}%
  \label{fig:EP4}%
\end{figure}

However, the behaviour changes completely for simultaneous perturbations in
$\tilde{\gamma}$ and $\tilde{\varepsilon}$. The eigenvalues $\mu_\pm$
\eref{eq:mu_merge} of the matrix $H$ with constant parameters $g=0.02$ and
$\varepsilon=0.0002\rmj$, and the parameter $\gamma$ along the path
$\gamma=\gamma_0v+0.0004\rme^{\,\rmj\vartheta}$ are shown in
figures~\ref{fig:EP4}\pt(a) and \pt(b) and compared to the numerical results of
the GPE.
The chemical potential is shifted by the mean value of all four states. The
solid lines for the numerically exact results of the GPE agree at least
qualitatively with the dashed lines indicating the eigenvalues of the matrix
model. In both the matrix model and the GPE the four eigenvalues move on a
single path in such a way that they permute after one cycle ($\vartheta=0\dots
2\pi$) in the parameter space, which provides clear evidence for the existence
of an EP4 in this system.

\section{Summary and outlook}
\label{sec:conclusion}
We have generalized the concept of linear $\PT$ symmetry to nonlinear
Gross-Pitaevskii-like systems of the type $\hat{H}_\mrm{lin}\psi+f(\psi)\psi
=\rmi\dot{\psi}$. If the spectrum is non-degenerate, the linear part
$\hat{H}_\mrm{lin}$ is $\PT$-symmetric and the function $f(\psi)$ fulfils the
two conditions,
\begin{eqnarray*}
  f(\rme^{\rmi\chi}\psi) &=& f(\psi), \quad \chi\in\mathbb{R},\\
  \PT f(\psi) &=& f(\PT\psi),
\end{eqnarray*}
the usual properties of linear $\PT$ symmetry hold, in particular
$\PT$-symmetric states and real eigenvalues coincide. These conditions imply
that $\PT$-symmetric wave functions render the nonlinear part and thus the
Hamiltonian $\PT$-symmetric.

The eigenvalue structure of one special system fulfilling the above-named
conditions was studied, namely a BEC with contact interaction in a
$\PT$-symmetric double well. The eigenvalue structure shows a tangent and a
pitchfork bifurcation at which the number of solutions changes. For further
investigation of the bifurcations an analytic continuation to bicomplex numbers
was introduced by complexification of the real and imaginary parts of the wave
functions. It was shown that this continuation leads to a conserved number of
solutions. Additionally it dictates new symmetries to the system and completes
the eigenvalue picture. The tangent bifurcation at which the two
$\PT$-symmetric solutions coalesce was found to be a second-order exceptional
point (EP2). At the pitchfork bifurcation the two $\PT$-broken solutions and
one $\PT$-symmetric solution coalesce. Due to studies in a $\PT$-symmetric
double-delta-trap potential~\cite{Heiss13a} we expect a third-order exceptional
point (EP3) at this bifurcation. To see the true nature of this exceptional
point, perturbations breaking the symmetry of the real part of the potential
were necessary. Using such perturbations revealed that this exceptional point
is indeed an EP3.

We derived a linear $4\times4$ matrix model which eigenvalues coincide with the
solutions of the non-Hermitian Bose-Hubbard model in mean-field limit by Graefe
\etal~\cite{Graefe12b,Graefe10}. It was shown that the eigenvalues of the
matrix model are in excellent agreement with the numerical exact solutions of
the GPE equation in the spatially extended double-well potential. Also both
the EP2 at the tangent bifurcation and the EP3 at the pitchfork bifurcation are
described by the matrix model. In the limit of vanishing interaction the matrix
model has the Jordan block of an EP4. On an appropriate path the permutation of
four states was observed in the matrix model and in the numerical results, thus
providing strong evidence for the existence also of an EP4.

Since we showed that $\PT$-symmetric effects should be observable for a certain
class of interaction types including the long-range dipolar interaction, such
investigations are a starting point for future work as well as for calculations
in more complex geometries.

\appendix
\section*{Appendix. Derivation of the linear matrix model}
\setcounter{section}{1}
\label{sec:appendix}
We want to construct a linear matrix model for $\PT$-symmetric
condensates with asymmetric perturbation where the eigenvalues of a
$4\times 4$ matrix $H$ agree with the chemical potential $\mu$ in
\eref{eq:mu_model} with $\bi{s}=(s_x,s_y,s_z)$ the fixed points of the
nonlinear and non-Hermitian Bloch equations \eref{eq:bloch} introduced
in \cite{Graefe10}.
(We omit the tildes on the parameters $\mu$, $\gamma$, $g$, and
$\varepsilon$ in the appendix.)
The Bloch equations yield
\begin{equation}
 \mu = g - 2\rmi \gamma s_z + \frac{\varepsilon}{2s_z}
\label{eq:mu}
\end{equation}
with $s_z$ the roots of the polynomial
\begin{equation}
 s_z^4 + \underbrace{\frac{g\varepsilon}{g^2+\gamma^2}}_{a_3} s_z^3
  + \underbrace{\frac{\varepsilon^2+v^2-g^2-\gamma^2}{g^2+\gamma^2}}_{a_2} s_z^2
  + \underbrace{\frac{-g\varepsilon}{4(g^2+\gamma^2)}}_{a_1} s_z
  + \underbrace{\frac{-\varepsilon^2}{16(g^2+\gamma^2)}}_{a_0} = 0 \; ,
\end{equation}
and
\begin{eqnarray*}
 a_3 &=& -(s_{z_1}+s_{z_2}+s_{z_3}+s_{z_4}) \; ,\quad
 a_2 = \sum_{i<j} s_{z_i} s_{z_j} \; ,\\
 a_1 &=& -\sum_{i<j<k} s_{z_i} s_{z_j} s_{z_k} \; ,\quad
 a_0 = s_{z_1} s_{z_2} s_{z_3} s_{z_4} \; .
\end{eqnarray*}
The decisive step is now the construction of the characteristic polynomial
\begin{equation}
 \chi(\mu) = \mu^4 + c_3 \mu^3 + c_2 \mu^2 + c_1 \mu + c_0
\label{eq:chi}
\end{equation}
of the matrix $H$.
The eigenvalues of $H$ coincide with the solutions $\mu_1$ to $\mu_4$
of \eref{eq:mu} when the conditions
\begin{eqnarray*}
 c_3 &=& -{\rm Tr}\, H = -(\mu_1+\mu_2+\mu_3+\mu_4) \; ,\quad
 c_2 = \sum_{i<j} \mu_i \mu_j \; ,\\
 c_1 &=& -\sum_{i<j<k} \mu_i \mu_j \mu_k \; ,\quad
 c_0 = \det H = \mu_1 \mu_2 \mu_3 \mu_4
\end{eqnarray*}
are fulfilled.
For the coefficient $c_3= -(\mu_1+\mu_2+\mu_3+\mu_4)$ we obtain
\begin{eqnarray*}
 c_3 &=& -4g + 2\rmi \gamma(s_{z_1}+s_{z_2}+s_{z_3}+s_{z_4}) - \frac{\varepsilon}{2}
   \left(\frac{1}{s_{z_1}}+\frac{1}{s_{z_2}}+\frac{1}{s_{z_3}}
    +\frac{1}{s_{z_4}}\right) \\
 &\Rightarrow&
 s_{z_1} s_{z_2} s_{z_3} s_{z_4} c_3 = a_0 c_3
 = -2 a_0 (2g + \rmi a_3 \gamma) + \frac{\varepsilon}{2} a_1 \; ,
\end{eqnarray*}
resulting in
\begin{equation}
\label{eq:c3}
 c_3 = -2 (2g + \rmi a_3 \gamma) + \frac{\varepsilon}{2} \frac{a_1}{a_0}
 = -2g - \frac{2\rmi g \gamma\varepsilon}{g^2+\gamma^2} \; .
\end{equation}
The coefficients $c_0$ to $c_2$ are derived in a similar way, though
more tediously, and read
\begin{eqnarray}
\label{eq:c2}
 c_2 &=& 2(\gamma^2-v^2) + g^2 + \frac{g^2v^2}{g^2+\gamma^2}
         + \frac{2\rmi \gamma(3g^2+2\gamma^2)}{g^2+\gamma^2}\varepsilon
         - \frac{g^2+2\gamma^2}{g^2+\gamma^2}\varepsilon^2 \; ,\\
\label{eq:c1}
 c_1 &=& -2g(\gamma^2-v^2) - \frac{2\rmi g
   \gamma(3\gamma^2+g^2-v^2)}{g^2+\gamma^2}\varepsilon \nonumber \\
         &+& \frac{6g \gamma^2}{g^2+\gamma^2}\varepsilon^2
         + \frac{2\rmi g \gamma}{g^2+\gamma^2}\varepsilon^3 \; ,\\
\label{eq:c0}
 c_0 &=& (\gamma^2-v^2)\frac{(g^2+\gamma^2)^2-\gamma^2v^2}{g^2+\gamma^2}
         + 4\rmi \gamma(\gamma^2-v^2)\varepsilon \nonumber \\
     &-& \frac{2\gamma^2(g^2+3\gamma^2-v^2)}{g^2+\gamma^2}\varepsilon^2
         - \frac{4\rmi \gamma^3}{g^2+\gamma^2}\varepsilon^3
         + \frac{\gamma^2}{g^2+\gamma^2}\varepsilon^4 \; .
\end{eqnarray}
It can now be checked directly that the characteristic polynomial of
the matrix $H$ given in \eref{eq:H_model} exactly coincides with
\eref{eq:chi}, and thus the eigenvalues of $H$ agree with the solution
of the nonlinear model of \cite{Graefe10}.

\end{document}